\documentclass[preprint,aps,12pt,notitlepage,nofootinbib,tightenlines]{revtex4}
\usepackage{amsmath}
\usepackage{caption}
\usepackage{enumerate}
\usepackage{bm}
\usepackage{times}
\usepackage{braket}
\usepackage{color}
\usepackage{epsfig}
\usepackage{slashed}
\usepackage{hyperref}
\usepackage{multirow}
\usepackage{booktabs}
\usepackage{array}
\usepackage{float}
\newcommand{\beq}{\begin{eqnarray}}
\newcommand{\eeq}{\end{eqnarray}}
\newcommand{\be}{\begin{equation}\begin{aligned}}
\newcommand{\ee}{\end{aligned}\end{equation}}

\definecolor{Red}{rgb}{1.,0.,0.}

\definecolor{Blue}{rgb}{0.,0.,1.}

\definecolor{nicered}{rgb}{0.7,0.1,0.1}
\definecolor{nicegreen}{rgb}{0.1,0.5,0.1}
\def\lsim{ {\ \lower-1.2pt\vbox{\hbox{\rlap{$<$}\lower6pt\vbox{\hbox{$\sim$}}}}\ } }
\def\gsim{ {\ \lower-1.2pt\vbox{\hbox{\rlap{$>$}\lower6pt\vbox{\hbox{$\sim$}}}}\ } }

\hypersetup{colorlinks,citecolor=nicegreen,linkcolor=nicered}
\begin{document}
\title{Searching for vectorlike $T$ quarks in the $T\to tZ$ channel at a future muon-proton collider}
\author{Yin-Hao Gao$^{1}$, Yao-Bei Liu$^{1,2}$\footnote{E-mail: liuyaobei@hist.edu.cn}}

\affiliation{1. Henan Institute of Science and Technology, Xinxiang 453003, China\\
2. School of Electro-Mechanical Engineering, Zhongyuan University of Science and Technology, Xuchang 461000, China\\}
\begin{abstract}
We investigate the discovery potential of a singly produced vector-like top quark $T$ in the $T\to tZ$ decay channel at future muon-proton ($\mu p$) colliders, considering three benchmark center-of-mass energies: $\sqrt{s}=5.29$, $6.48$, and $9.16$~TeV. The baseline signal final state consists of a semileptonic top decay $t\to bW\to b\ell\nu$ and a hadronic $Z\to q\bar{q}$ decay, where the collimated quark pair from the $Z$ boson is reconstructed as a single large-radius fat jet $J$. Full detector-level Monte Carlo simulations are performed for both the signal process and the dominant Standard Model backgrounds. We derive the $5\sigma$ discovery contours and the $95\%$ confidence level exclusion limits in the two-dimensional parameter space spanned by the VLQ mass $m_T$ and the effective coupling $g^*$. A comparison with the $T\to Wb$ channel and with searches at other collider facilities reveals that the $T\to tZ$ decay mode provides complementary sensitivity and offers unique advantages for probing high-mass vector-like $T$ quarks in the boosted jet topology.
\end{abstract}

\maketitle

\newpage
\section{Introduction}
Vector-like quarks (VLQs) with masses at the TeV scale are generally predicted in a variety of extensions of the Standard Model~(SM), such as little Higgs models~\cite{ArkaniHamed:2002qy,Schmaltz:2005ky}, composite Higgs models~\cite{Agashe:2004rs,Contino:2006qr}, two-Higgs-doublet models~\cite{Benbrik:2022kpo,Arhrib:2024tzm,Arhrib:2024dou,Arhrib:2024nbj,Benbrik:2024hsf,Arhrib:2024mbq}, and other extended models~\cite{He:1999vp,Wang:2013jwa,He:2001fz,He:2014ora}. A common feature of these new particles is that the left- and right-handed chiral components transform in the same way under the electroweak (EW) symmetry group of the SM~\cite{Aguilar-Saavedra:2013qpa}. Unlike for chiral quarks, bare mass terms of VLQs are gauge invariant and therefore they can avoid the stringent constraints from Higgs boson data~\cite{He:2001tp,Chen:2012wz}. Moreover, VLQs have the potential to stabilize the EW vacuum~\cite{Xiao:2014kba,Cingiloglu:2023ylm}, address the so-called Cabibbo-Kobayashi-Maskawa (CKM) unitarity problem~\cite{Cheung:2020vqm,Crivellin:2022rhw,Belfatto:2021jhf,Branco:2021vhs,Botella:2021uxz}, and may also provide explanations for various experimental anomalies, such as the $W$-mass anomaly~\cite{Cao:2022mif,Crivellin:2022fdf,He:2022zjz,Branco:2022gja,Abouabid:2023mbu}. VLQs can give rise to a rich variety of phenomena at the Large Hadron Collider (LHC) and future high-energy colliders~(see, e.g.,~\cite{DeSimone:2012fs,Buchkremer:2013bha,Aguilar-Saavedra:2009xmz,Mrazek:2009yu,Dissertori:2010ug,Atre:2011ae,Cacciapaglia:2011fx,Cacciapaglia:2012dd,Gopalakrishna:2013hua,Matsedonskyi:2014mna,Backovic:2014uma,
Chen:2016yfv,Fuks:2016ftf,Liu:2016jho,Aguilar-Saavedra:2017giu,Cui:2022hjg,Xie:2019gya,Benbrik:2019zdp,Aguilar-Saavedra:2019ghg,Belyaev:2021zgq,
Bhardwaj:2022nko,Bhardwaj:2022wfz,Verma:2022nyd,Bardhan:2022sif,Alves:2023ufm,Canbay:2023vmj,Belyaev:2023yym,Liu:2024hvp,
Shang:2024wwy,Cetinkaya:2020yjf,Zhang:2024ncj,Yang:2024aav,Mandal:2026bit,Shang:2026smr,Shang:2026one}).

The VLQ $T$-state (denoted $T$ hereafter) carries an electric charge of $+2e/3$ and can appear in various weak-isospin multiplets. For a singlet VLQ-$T$, three decay modes are possible: $T\to bW$, $tZ$, and $th$ (unless an extended Higgs sector is present~\cite{Aguilar-Saavedra:2017giu,Benbrik:2019zdp}). In the high-mass limit, the branching ratios (BRs) satisfy BR$(T\to th)\approx {\rm BR}(T\to tZ)\approx \frac{1}{2}{\rm BR}(T\to Wb)$. To date, the ATLAS and CMS Collaborations have conducted extensive searches for pair-produced VLQs, obtaining mass constraints at $95\%$ confidence level~(CL)~\cite{ATLAS:2024gyc,ATLAS:2022hnn,CMS:2022fck,Aaboud:2018wxv,Aaboud:2018xpj,Aaboud:2018uek,Aaboud:2018ifs,Sirunyan:2018qau,Sirunyan:2018omb,Aaboud:2018pii,CMS:2019eqb,Buckley:2020wzk}. For instance, the ATLAS Collaboration has set a lower bound of about $1.36$ TeV on the mass of a singlet VLQ-$T$ using $140~\text{fb}^{-1}$ of data~\cite{ATLAS:2024gyc}. The CMS Collaboration has excluded a singlet VLQ-$T$ mass below $1.46$ TeV at $95\%$ CL using $138~\text{fb}^{-1}$ of $pp$ collision data in leptonic final states~\cite{CMS:2022fck}. In addition, VLQ-$T$ can be singly produced at the LHC via EW interactions, and these processes are highly sensitive to the couplings between VLQs and SM quarks~\cite{Moretti:2016gkr,Carvalho:2018jkq}. Recent searches by the ATLAS and CMS Collaborations have set limits on VLQ masses and couplings using Run 2 data~\cite{ATLAS:2022ozf,ATLAS:2023pja,ATLAS:2023bfh,ATLAS:2024xdc,ATLAS:2024kgp,CMS:2023agg,CMS:2024qdd,CMS:2024bni}.

In this work, we consider a muon-proton collider with multi-TeV beam energies~\cite{Shiltsev:1997pv,Ginzburg:1998yw,Cheung:1997rg,Cheung:1999wy,Carena:2000su,Kaya:2018kyt,Kaya:2019ecf,Ketenoglu:2022fzo,Dagli:2022idi,Kaya:2022wrc,Akturk:2024evo}. Compared to the LHC, BSM studies at such a machine typically benefit from significantly reduced QCD backgrounds. Recent phenomenological studies for a future $\mu p$ collider can be found in Refs.~\cite{Caliskan:2017meb,Acar:2017eli,Caliskan:2018vep,Caliskan:2018,Alici:2019asv,Ozansoy:2019rmu,Spor:2020rig,Cheung:2021iev,Aydin:2021iky,Gurkanli:2024tfo,Alici:2024eez,Benbrik:2026zjv}. Very recently, the authors of Ref.~\cite{Han:2025itd} systematically investigated the single production of vector-like quarks $T$ and $Y$ decaying into $Wb$ at a future $\mu p$ collider, considering both the leptonic and hadronic decay channels of the $W$ boson. Their results demonstrate that the hadronic channel, in which the $W$ decay products merge into a fat jet, enables a $5\sigma$ discovery up to $m_T = 3750$ GeV and $m_Y = 4100$ GeV at $\sqrt{s}=9.16$ TeV with an integrated luminosity of $100~\text{fb}^{-1}$, thereby significantly extending the reach beyond LHC capabilities. Nevertheless, the complementary decay mode $T \to tZ$ has not yet been explored in the context of a $\mu p$ collider. In the high-mass limit, the branching ratio of $T \to tZ$ is approximately half that of $T \to Wb$, i.e., $\mathrm{BR}(T\to tZ) \approx \frac{1}{2}\mathrm{BR}(T\to Wb)$, making this channel a promising and complementary avenue for VLQ searches. Moreover, the distinctive final-state topology, featuring a leptonically decaying top quark and a hadronically decaying $Z$ boson (reconstructed as a fat jet), provides unique kinematic handles for suppressing SM backgrounds.

Motivated by the above considerations, we focus on the observability of singly produced vector-like $T$ quarks at a future $\mu p$ collider via the $T\to tZ$ decay channel. The top quark decays semileptonically as $t\to bW\to b\ell\nu$, and the $Z$ boson decays hadronically into $q\bar{q}$, with the collimated quark pair reconstructed as a single fat jet $J$. We perform full detector-level Monte Carlo simulations for three benchmark center-of-mass~(c.m.) energies $\sqrt{s}=5.29$, $6.48$, and $9.16$ TeV, optimize the kinematic event selection cuts, and extract the $5\sigma$ discovery contours and $95\%$ CL exclusion limits in the two-dimensional $(g^{*}, m_T)$ parameter plane. We further provide a dedicated comparison with the $T\to Wb$ channel to demonstrate the complementary discovery potential of these two decay signatures.

The remainder of this paper is organized as follows. In Sec.~II, we introduce the simplified effective model of the vector-like quark $T$ and calculate its single-production cross sections at the $\mu p$ collider for three different c.m. energies. In Sec.~III, we investigate the observability of the signal through the full decay chain $T\to tZ\to bWZ\to b\ell\nu J$, which gives rise to a final state with missing transverse momentum from neutrinos. We perform full detector-level Monte Carlo simulations, construct optimized kinematic selection cuts, and present the resulting $5\sigma$ discovery and $95\%$ CL exclusion reaches. Section~IV draws the main conclusions and provides supplementary discussions on alternative decay topologies.

\section{Simplified model and single $T$ production at a $\mu p$ collider}

In this work, we adopt a singlet vector-like quark $T$ under $SU(2)_L$ as our benchmark particle. This minimal particle content is well motivated by numerous ultraviolet (UV)-complete BSM frameworks and has been widely used as a standard reference scenario for collider searches targeting vector-like quarks. This simplified setup allows us to perform a largely model-independent leading-order analysis, without introducing extra free parameters associated with additional exotic particles.

Following the conventions of Ref.~\cite{Buchkremer:2013bha}, the general effective Lagrangian describing the singlet VLQ-$T$ is written as
\beq
\hspace*{-0.15cm}
{\cal L}_{\rm T} =\frac{gg^{\ast}}{2}\left\{\frac{1}{\sqrt{2}}\left[\bar{T}_{L}W_{\mu}^{+}
    \gamma^{\mu} b_{L}\right]+
    \frac{1}{2c_W}\left[\bar{T}_{L} Z_{\mu} \gamma^{\mu} t_{L}\right]
    - \frac{m_{T}}{2m_{W}}\left[\bar{T}_{R}ht_{L}\right] -\frac{m_{t}}{2m_{W}} \left[\bar{T}_{L}ht_{R}\right]\right\}+ {\rm H.c.},
  \label{TsingletVL}
\eeq
where $g$ is the $SU(2)_L$ gauge coupling constant, $c_W=\cos\theta_W$, and $\theta_W$ is the Weinberg angle. Thus, only two model parameters remain: the VLQ-$T$ mass $m_T$ and the effective coupling $g^{\ast}$, which is defined in units of the SM gauge coupling. As mentioned, the singlet VLQ-$T$ has three possible decay modes: $T\to bW$, $tZ$, and $th$. For $m_T \geq 1$ TeV, the branching ratios satisfy ${\rm BR}(T\to th)\approx {\rm BR}(T\to tZ)\approx \frac{1}{2}{\rm BR}(T\to Wb)$, as expected from the Goldstone boson equivalence theorem~\cite{He:1992nga,He:1993yd,He:1994br,He:1996rb,He:1996cm}. For $m_T = 1.5$ (2.0) TeV, $g^{\ast}$ is constrained to be smaller than about $0.37$ ($0.59$)~\cite{Benbrik:2024fku,Arhrib:2026coy}. In this work, we adopt a phenomenological upper limit of $g^{\ast}\leq 0.5$ for $m_T \geq 1.5$ TeV.

The VLQ-$T$ can be singly produced in $\mu p$ collisions through $Wb$ fusion and subsequently decays via $T\to tZ$. Leading-order (LO) cross sections are computed with MadGraph5-aMC@NLO~\cite{Alwall:2014hca}, using the default NNPDF23L01 parton distribution functions (PDFs)~\cite{Ball:2014uwa} together with the standard renormalization and factorization scales. Three representative collision setups are considered, corresponding to c.m. energies $\sqrt{s}=5.29$, $6.48$, and $9.16$~TeV, with a fixed proton beam energy $E_p=7$ TeV and muon beam energies $E_\mu=1.0$, $1.5$, and $3.0$ TeV, respectively. In our leading-order simulations, the muon beam is assumed to be unpolarized.

\begin{figure}[thb]
\begin{center}
\vspace{-0.5cm}
\centerline{\epsfxsize=10cm \epsffile{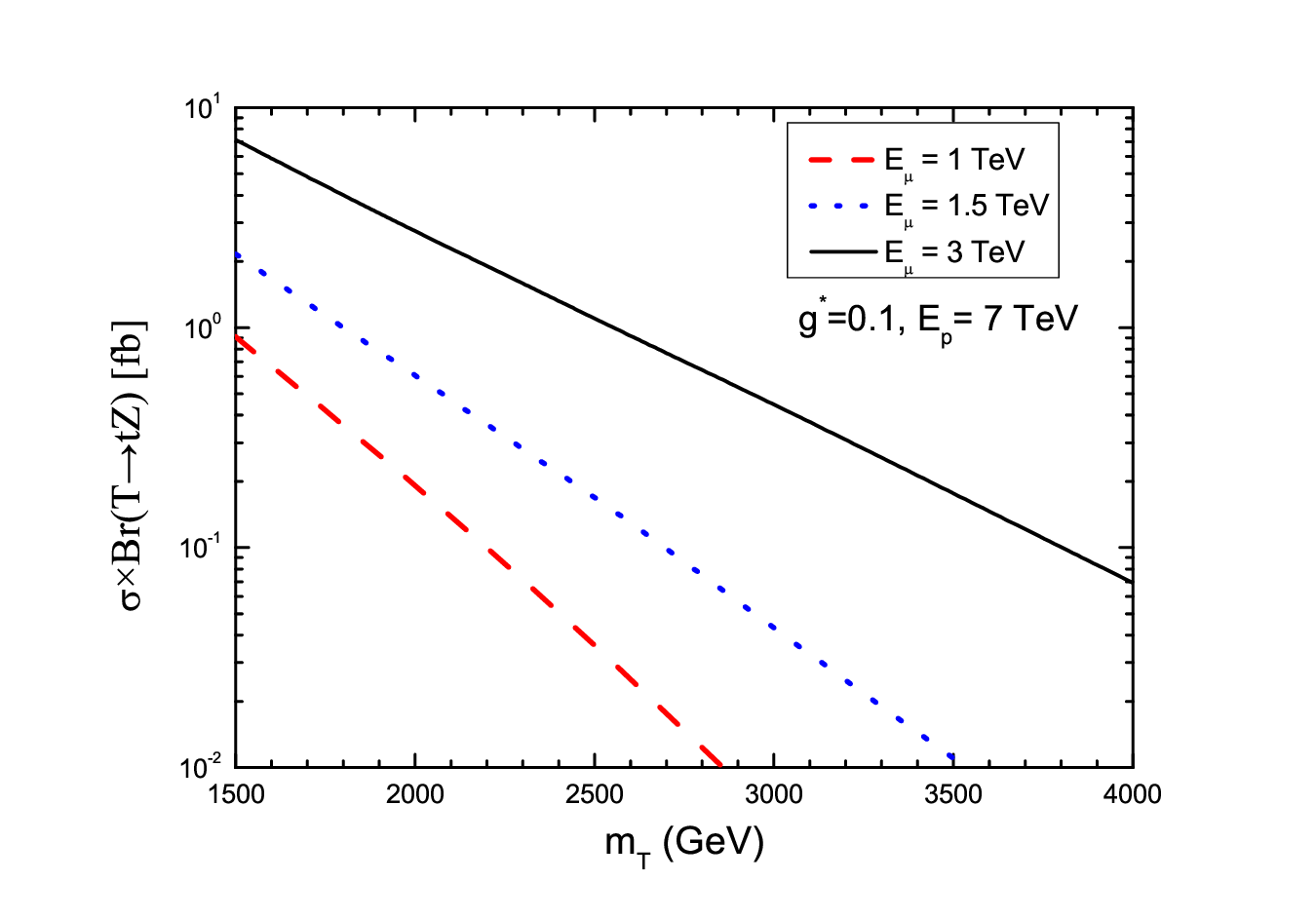}}
\caption{Effective cross section $\sigma\times{\rm BR}(T\to tZ)$ as a function of the VLQ-$T$ mass for $g^{*}=0.1$ at a $\mu p$ collider with three different c.m. energies.}
\label{cross}
\end{center}
\end{figure}

In Fig.~\ref{cross}, we display the effective cross sections $\sigma\times{\rm BR}(T\to tZ)$ as a function of $m_T$ for the benchmark coupling $g^{*}=0.1$, where all conjugate production processes are summed. The cross section decreases gradually with increasing $m_T$, driven by the shrinking available phase space. For $g^{*}=0.1$ and $m_T=2$ TeV, the corresponding cross sections are $0.19$, $0.61$, and $2.74$ fb at $\sqrt{s}=5.29$, $6.48$, and $9.16$~TeV, respectively. As anticipated, the single-production cross section scales quadratically with the effective coupling, i.e., $\sigma \propto (g^{*})^2$.

\section{Collider simulation and analysis}

\begin{figure}[h]
\centering
\includegraphics[width = 16cm ]{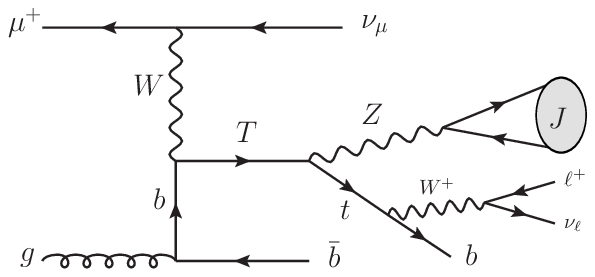}
\vspace{-17cm}
\caption{Representative tree-level Feynman diagram for single VLQ-$T$ production at a $\mu p$ collider with the decay cascade $T\to tZ$. The top decays semileptonically via $t\to W^{+}b\to \ell^{+}\nu_{\ell} b$, and $Z\to q\bar{q}$ is reconstructed as a merged fat jet $J$.}
\label{fig:fey}
\end{figure}

We carry out Monte Carlo (MC) simulations for both signal and SM background events corresponding to the process
\[
\mu^{+}p \to \bar{\nu}_{\mu} T (\to tZ) \bar{b} \to \bar{\nu}_{\mu} \big(b \ell^{+}\nu_{\ell}\big) \big(q\bar{q}\big) \bar{b},
\]
where the intermediate $W^{+}$ boson decays leptonically as $W^{+}\to \ell^{+}\nu_{\ell}$ with $\ell=e,\mu$, and the $Z$ boson decays hadronically via $Z\to q\bar{q}$. The resulting final state consists of an isolated charged lepton, one $b$-tagged jet originating from top quark decay, a single fat jet $J$ reconstructed from the collimated products of the $Z\to q\bar{q}$ decay, and missing transverse momentum arising from undetected neutrinos. The kinematic features of all conjugate production processes are analogous, and we include their contributions in our event yields. A representative tree-level Feynman diagram for this signal process is shown in Fig.~\ref{fig:fey}.

The dominant irreducible SM backgrounds are categorized as follows:
\begin{itemize}
    \item $\mu p \to \nu_{\mu} t j$, where the top quark decays semileptonically via $t\to Wb\to b\ell\nu$;
    \item $\mu p \to \nu_{\mu} t Z j$, with semileptonic top decay $t\to Wb\to b\ell\nu$ and hadronic $Z\to q\bar{q}$ decay;
    \item $\mu p \to \nu_{\mu} W Z j$, featuring leptonic $W\to \ell\nu$ and hadronic $Z\to q\bar{q}$ decays.
\end{itemize}

In addition, we study reducible backgrounds dominated by QCD multi-jet processes, including $\mu p \to \nu_\mu jjj$ and $\mu p \to \nu_\mu b\bar{b}j$. Such backgrounds enter the signal region through two main fake sources: charged hadrons that fail isolation criteria and are misreconstructed as prompt light leptons, and fake missing transverse momentum $\slashed{E}_T$ arising from fluctuations in the jet energy scale. To suppress these QCD contributions, we apply a suite of tight kinematic vetoes throughout the analysis. A stringent lepton isolation working point largely eliminates fake leptons from hadronic activity, while complementary requirements on the number of $b$-tagged jets, a high minimum $\slashed{E}_T$ threshold, and a dedicated boosted $Z$-jet veto collectively further reduce the QCD background yields.

Taking the $\mu p \to \nu_\mu b\bar{b}j$ process as a concrete example, its inclusive cross section at a c.m. energy of $\sqrt{s}=5.29\,\text{TeV}$ is approximately $440\,\text{fb}$. After all analysis selections are imposed, the overall event acceptance efficiency is found to be below the $10^{-7}$ level, as estimated from full MC simulation. Although these reducible QCD backgrounds do not vanish entirely, their total contribution to the background budget lies far below the statistical uncertainty of the expected signal yield. For this reason, they are safely neglected in the subsequent signal extraction procedure.

Table~\ref{cross-sm} lists the cross sections times branching ratios for all considered irreducible backgrounds at the three c.m. energies.

\begin{table}[htb]
\centering
\caption{Cross sections times branching ratios $\sigma \times \mathcal{B}$ (unit: fb) of three SM backgrounds at three c.m. energies $\sqrt{s}=5.29,\,6.48,\,9.16$ TeV. \label{cross-sm}}
\vspace{0.8cm}
\begin{tabular}{p{3.0cm}<{\centering} p{2.5cm}<{\centering} p{2.5cm}<{\centering} p{2.0cm}<{\centering} p{2.0cm}<{\centering}}
\toprule[1.5pt]
Process & Decay Channel & $\sqrt{s}=5.29$ & $6.48$ & $9.16$  \\ \hline
$\mu p\to\nu_{\mu} tj$ & $t\to b\ell\nu$ & 12840 & 17800& 29930\\
$\mu p\to\nu_{\mu} t Z j$ & $Z\to q\bar{q}$ & 31.6 & 49.3 & 99.4 \\
$\mu p\to\nu_{\mu} W Z j$ & $W/Z\to \ell\nu/q\bar{q}$ & 18.3 & 26.6 & 48.2\\
\hline
\end{tabular}
\end{table}

Signal and background events are generated at LO with MadGraph5-aMC$@$NLO using the same PDFs. Parton showering and hadronization are performed with Pythia 8.3~\cite{pythia8}. Detector simulation is carried out with Delphes 3.4.2~\cite{deFavereau:2013fsa} using the standard LHeC detector card. Small-radius jets are reconstructed with the anti-$k_t$ algorithm~\cite{Cacciari:2008gp} with $R=0.4$. Fat jets are reconstructed with the Cambridge-Aachen algorithm~\cite{Dokshitzer:1997in,Wobisch:1998wt} with $R=0.8$. The $b$-tagging efficiency is $80\%$, with misidentification rates of $0.1\%$ for light-flavor jets and $5\%$ for charm jets. Analysis is performed with MadAnalysis 5~\cite{Conte:2012fm,Conte:2014zja}.

Basic object selection cuts are:
\[
p_{T}^{\ell/b/j} > 15~\text{GeV},\quad |\eta_\ell| < 2.5,\quad |\eta_{j/b}| < 5,\quad \Delta R_{xy} > 0.4, \quad \slashed{E}_T > 30~\text{GeV},
\]
where $p_{T}^{\ell/b/j}$ and $\eta_{\ell/b/j}$ are the transverse momentum and pseudorapidity of leptons, $b$-jets, and light jets. Here, $\Delta R(x,y)=\sqrt{\Delta\phi^{2}+\Delta\eta^{2}}$ is the separation in the pseudorapidity-azimuth plane between the objects $x$ and $y$, where $x,y=\ell, b, j$.

We consider three signal benchmark points with $g^{*}=0.1$: $T_{1500}$ ($m_T = 1500$ GeV), $T_{2000}$ ($m_T = 2000$ GeV), and $T_{2500}$ ($m_T = 2500$ GeV).

\begin{figure*}[htb]
\begin{center}
\centerline{\hspace{2.0cm}\epsfxsize=9cm\epsffile{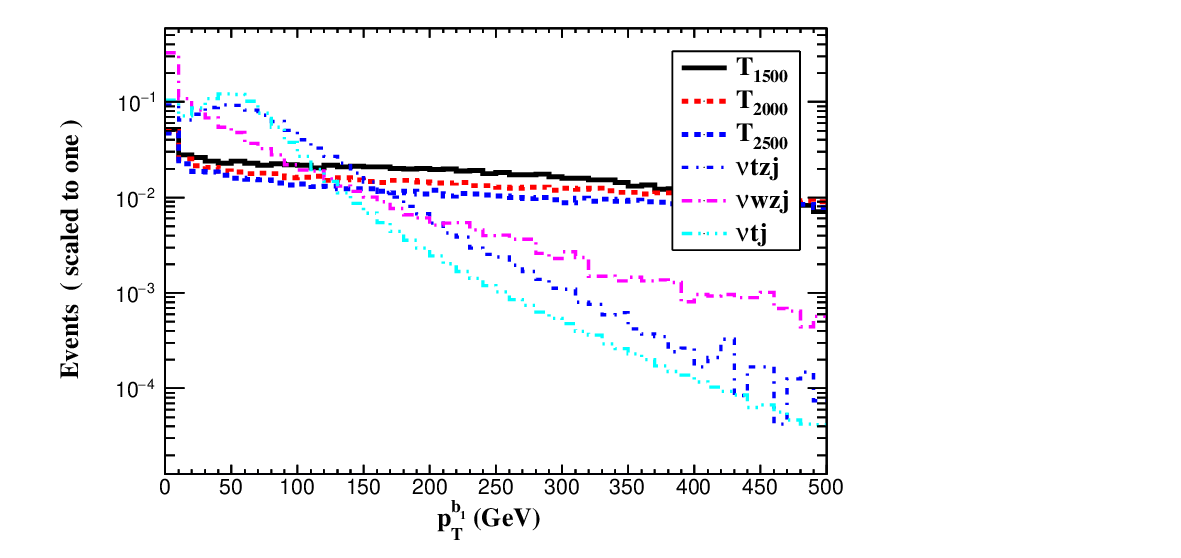}
\hspace{-2.0cm}\epsfxsize=9cm\epsffile{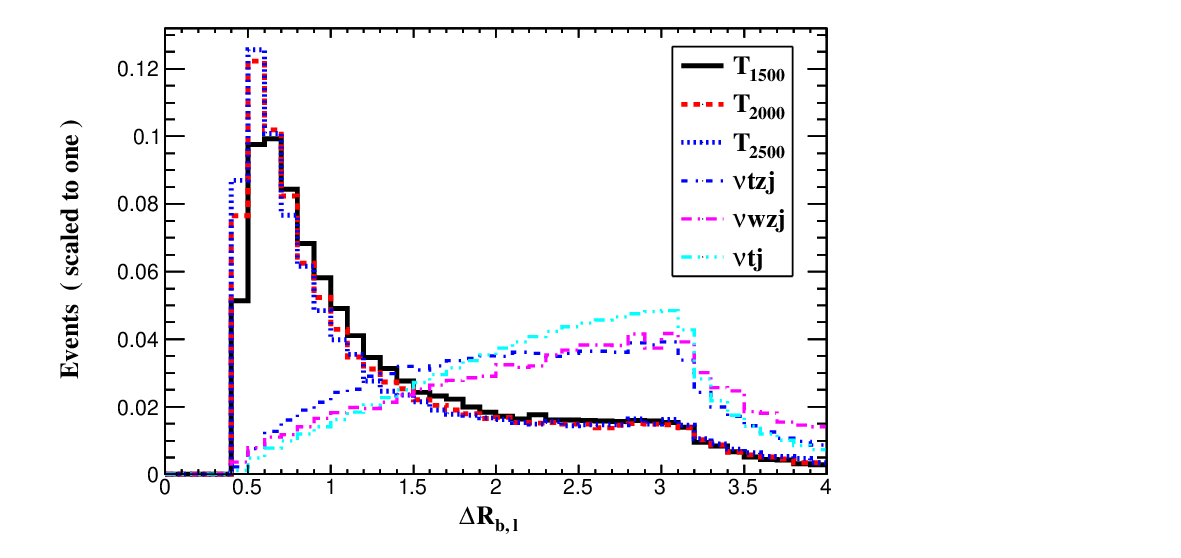}}
\centerline{\hspace{2.0cm}\epsfxsize=9cm\epsffile{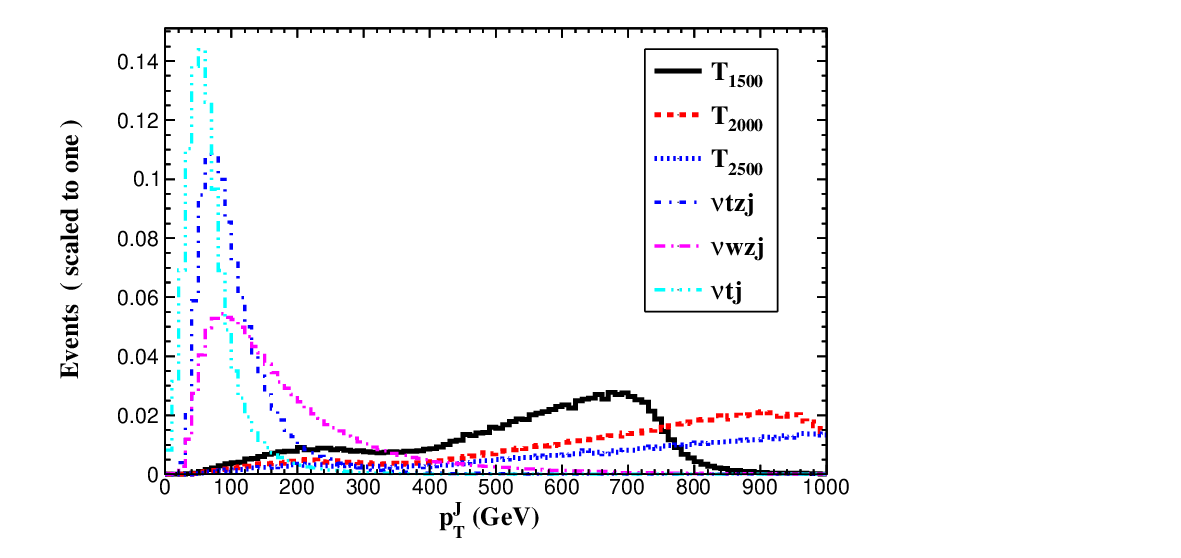}
\hspace{-2.0cm}\epsfxsize=9cm\epsffile{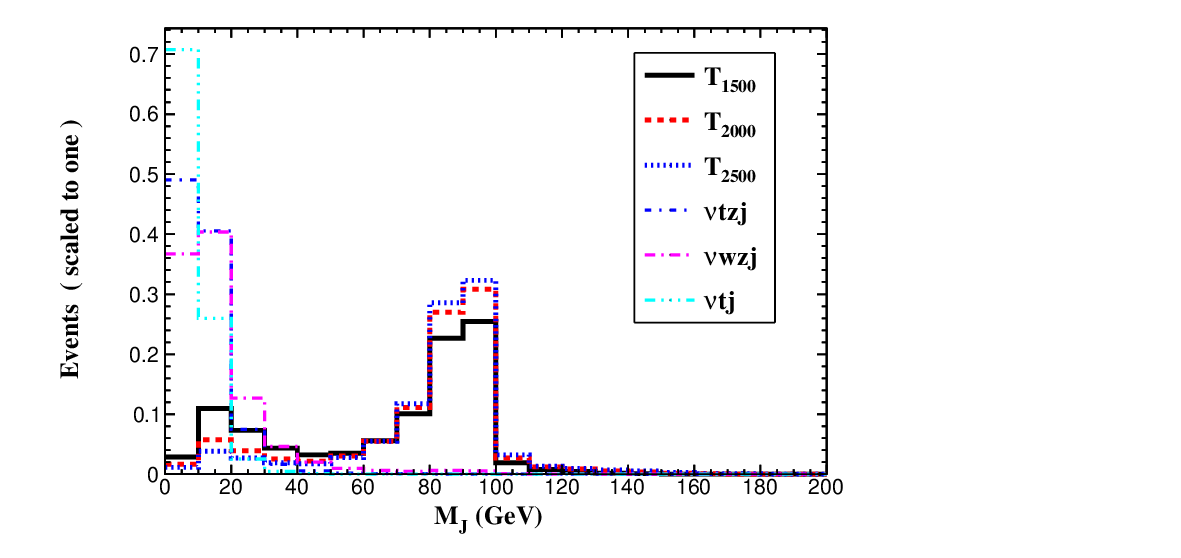}}
\centerline{\hspace{2.0cm}\epsfxsize=9cm\epsffile{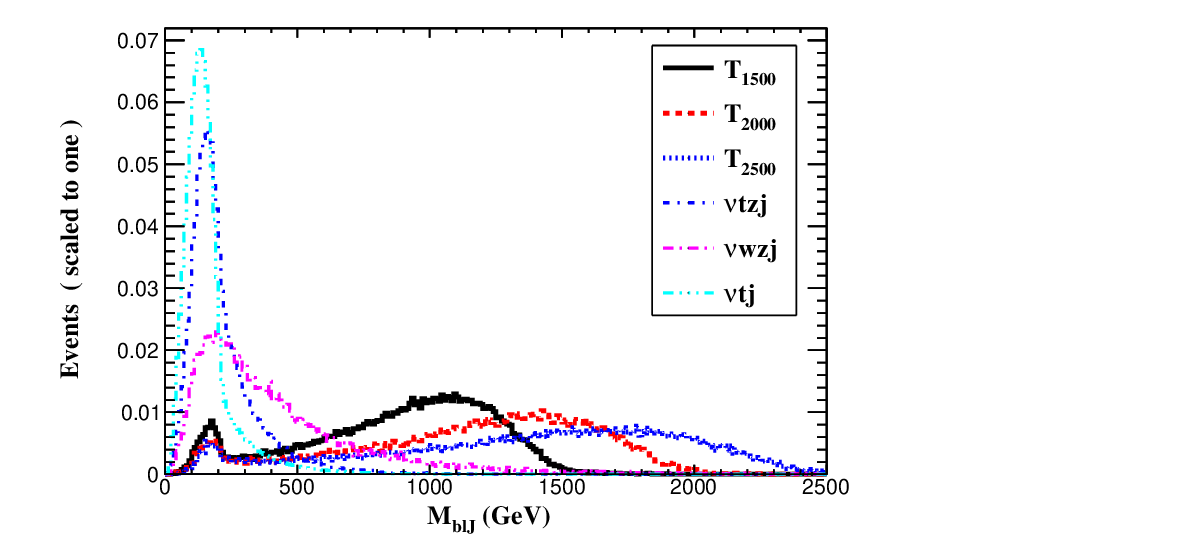}}
\caption{Normalized distributions of $p_T^{b}$, $\Delta R_{b,\ell}$, $p_T^{J}$, $M_{J}$, and $M_{b\ell J}$ for the three signal benchmarks with $m_T = 1500$ GeV (solid), $2000$ GeV (dashed), and $2500$ GeV (dotted), along with the SM backgrounds at $\sqrt{s}=5.29$~TeV.}
\label{fig3}
\end{center}
\end{figure*}

Figure~\ref{fig3} displays the normalized kinematic distributions for the signals and SM backgrounds at $\sqrt{s}=5.29$~TeV, showing the transverse momentum of the leading $b$-jet ($p_T^{b}$), the angular separation between the $b$-jet and the lepton ($\Delta R_{b,\ell}$), the fat jet transverse momentum ($p_T^J$), the fat jet invariant mass ($M_J$), and the invariant mass of the lepton-$b$-jet-fat jet system ($M_{b\ell J}$). Signal events exhibit harder $p_T$ spectra and a distinct peak in $M_J$ around the $Z$-boson mass.

Based on these distributions, we impose the following selection criteria:
\begin{itemize}
\item Cut 1: At least one fat jet with $p_T^J > 400$ GeV and $|M_J - m_Z| < 10$ GeV.
\item Cut 2: At least one isolated lepton and at least one $b$-tagged jet from the top decay ($N_b \geq 1$), with the leading $b$-jet satisfying $p_T^{b} > 150$ GeV and $\Delta R_{b,\ell} < 1.0$.
\item Cut 3: $M_{b\ell J} > 800$ GeV.
\end{itemize}

The kinematic distributions at $\sqrt{s}=6.48$ and $9.16$~TeV are found to be similar to those at $5.29$~TeV, with differences in the key shape variables below the per-mille level; they are therefore omitted for brevity. The same set of selection cuts is applied uniformly across all three c.m. energies, as the optimal cut values were found to depend only weakly on $\sqrt{s}$. The corresponding cut-flow tables are presented in Table~\ref{cutflows}. As shown, all SM backgrounds are suppressed very efficiently, while the signal efficiencies remain at the percent level after the full selection, which is substantially higher than those of the backgrounds. The total SM background cross sections after all cuts amount to $0.0078$ fb at $5.29$ TeV, $0.014$ fb at $6.48$ TeV, and $0.029$ fb at $9.16$ TeV.

\begin{table}[htb]
\centering
\small
\caption{Cut flow of cross sections (in fb) for signals and SM backgrounds in the $tZ$ channel with $g^{*}=0.1$ at different c.m. energies.}
\vspace{0.3cm}

\textbf{(a) $\sqrt{s}=5.29$ TeV}
\begin{tabular}{lccc|ccc}
\toprule[1.5pt]
Cuts & $T_{1500}$ & $T_{2000}$ & $T_{2500}$ & $\nu tj$ & $\nu tZj$ & $\nu WZj$ \\ \midrule
Basic & 0.0544 & 0.0091 & 0.0013 & 7961 & 21 & 13.4 \\
Cut 1 & 0.026 & 0.005 & 0.00074 & 0.28 & 0.023 & 0.155 \\
Cut 2 & 0.0091 & 0.002 & 0.00031 & 0.022 & 0.0052 & 0.00053 \\
Cut 3 & 0.009 & 0.0198 & 0.0003 & 0.00038 & 0.00367 & 0.00024 \\
Total eff. & 8.9\% & 9\% & 7.5\% & $3\times10^{-7}$ & $1.16\times10^{-4}$ & $1.32\times10^{-5}$ \\ \hline
\bottomrule
\end{tabular}

\vspace{0.5cm}

\textbf{(b) $\sqrt{s}=6.48$ TeV}
\begin{tabular}{lccc|ccc}
\toprule[1.5pt]
Cuts & $T_{1500}$ & $T_{2000}$ & $T_{2500}$ & $\nu tj$ & $\nu tZj$ & $\nu WZj$ \\ \midrule
Basic & 0.132 & 0.029 & 0.0061 & 11036 & 31 & 19.4 \\
Cut 1 & 0.062 & 0.016 & 0.0036 & 0.43 & 0.049 & 0.266\\
Cut 2 & 0.021 & 0.006 & 0.0014 & 0.03 & 0.0093 & 0.00079 \\
Cut 3 & 0.021 & 0.006 & 0.0014 & 0.0053 & 0.0083 & 0.00032 \\
Total eff. & 8.94\% & 8.94\% & 7.4\% & $3\times10^{-7}$ & $1.69\times10^{-4}$ & $1.19\times10^{-5}$ \\ \hline
\bottomrule
\end{tabular}

\vspace{0.5cm}

\textbf{(c) $\sqrt{s}=9.16$ TeV}
\begin{tabular}{lccc|ccc}
\toprule[1.5pt]
Cuts & $T_{1500}$ & $T_{2000}$ & $T_{2500}$ & $\nu tj$ & $\nu tZj$ & $\nu WZj$ \\ \midrule
Basic & 0.436 & 0.129 & 0.039 & 18557 & 53.7 & 34.2 \\
Cut 1 & 0.21 & 0.07 & 0.022 & 0.89 & 0.18 & 0.67 \\
Cut 2 & 0.069 & 0.027 & 0.0089 & 0.051 & 0.026 & 0.0014 \\
Cut 3 & 0.068 & 0.027 & 0.0089 & 0.009 & 0.019 & 0.00057 \\
Total eff. & 8.55\% & 8.8\% & 7.3\% & $3\times10^{-7}$ & $1.92\times10^{-4}$ & $1.18\times10^{-5}$ \\ \hline
\bottomrule
\end{tabular}
\label{cutflows}
\end{table}

We estimate the expected discovery ($\mathcal{Z}_{\text{dis}}$) and exclusion ($\mathcal{Z}_{\text{exc}}$) significances using the formulas from Ref.~\cite{Cowan:2010js}:
\be
\mathcal{Z}_\text{dis} =
  \sqrt{2\left[(s+b)\ln\left(\frac{(s+b)(1+\delta_{sys}^2 b)}{b+\delta_{sys}^2 b(s+b)}\right) -
  \frac{1}{\delta_{sys}^2 }\ln\left(1+\frac{\delta_{sys}^2 s}{1+\delta_{sys}^2 b}\right)\right]},
\ee
\be
\mathcal{Z}_\text{exc} =\sqrt{2\left[s-b\ln\left(\frac{b+s+x}{2b}\right)
  - \frac{1}{\delta_{sys}^2 }\ln\left(\frac{b-s+x}{2b}\right)\right] -
  \left(b+s-x\right)\left(1+\frac{1}{\delta_{sys}^2 b}\right)},
\ee
with $x=\sqrt{(s+b)^2- 4 \delta_{sys}^2 s b^2/(1+\delta_{sys}^2 b)}$, where $s$ and $b$ denote the numbers of signal and background events, respectively, and $\delta_{sys}$ represents the systematic uncertainty. The integrated luminosity is assumed to be $1000~\text{fb}^{-1}$. We consider a systematic uncertainty of $\delta_{sys} = 20\%$ as our default benchmark. This value is chosen to be conservative and is representative of the combined systematic effects expected at a future multi-TeV collider, including uncertainties in the jet energy scale, $b$-tagging efficiency, lepton identification, missing transverse momentum resolution, and luminosity determination. To illustrate the impact of this assumption, we also present results for the idealized scenario with no systematic uncertainty ($\delta_{sys} = 0$).

\begin{figure}[htb]
\begin{center}
\centerline{\epsfxsize=7.5cm \epsffile{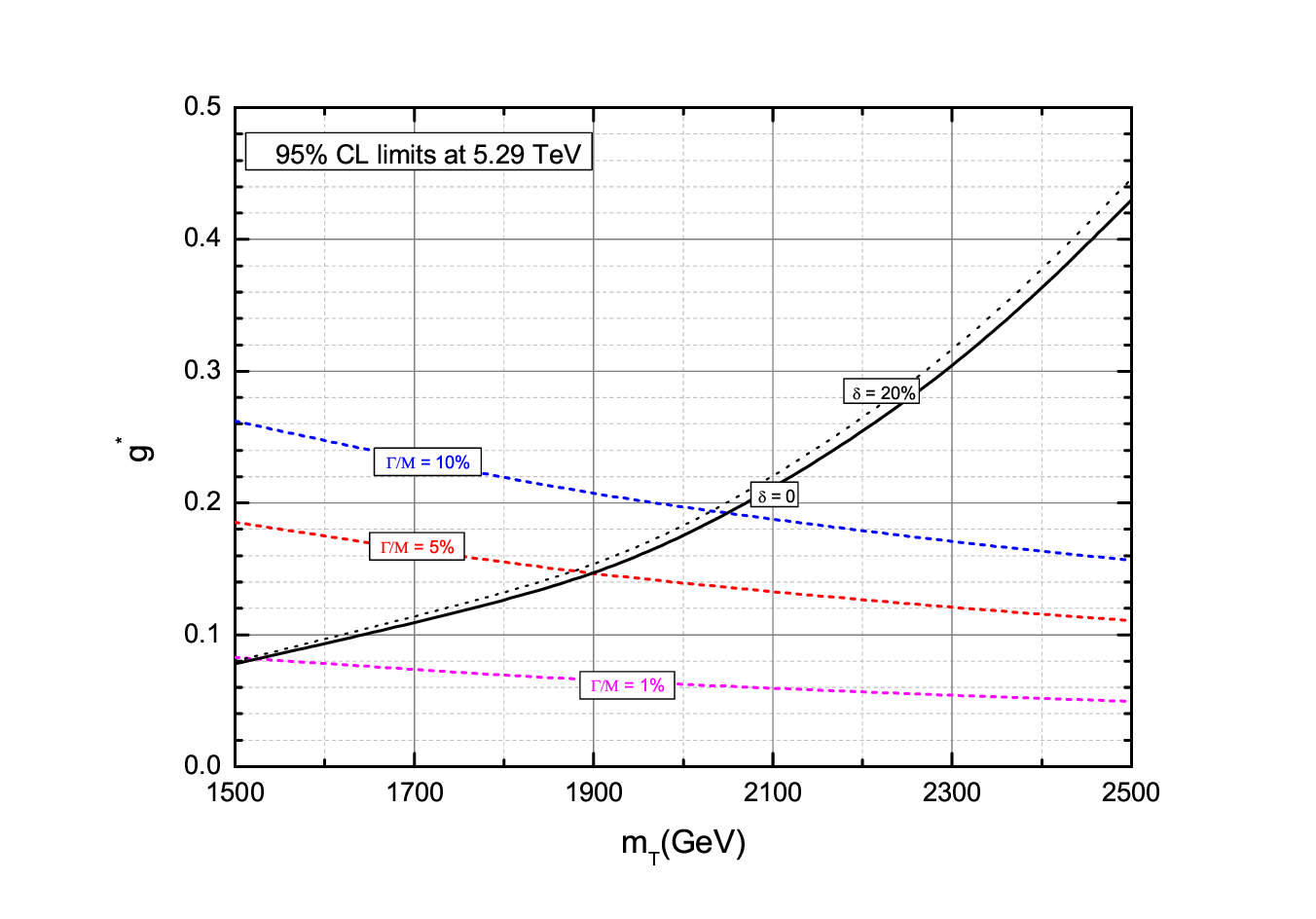}\epsfxsize=7.5cm \epsffile{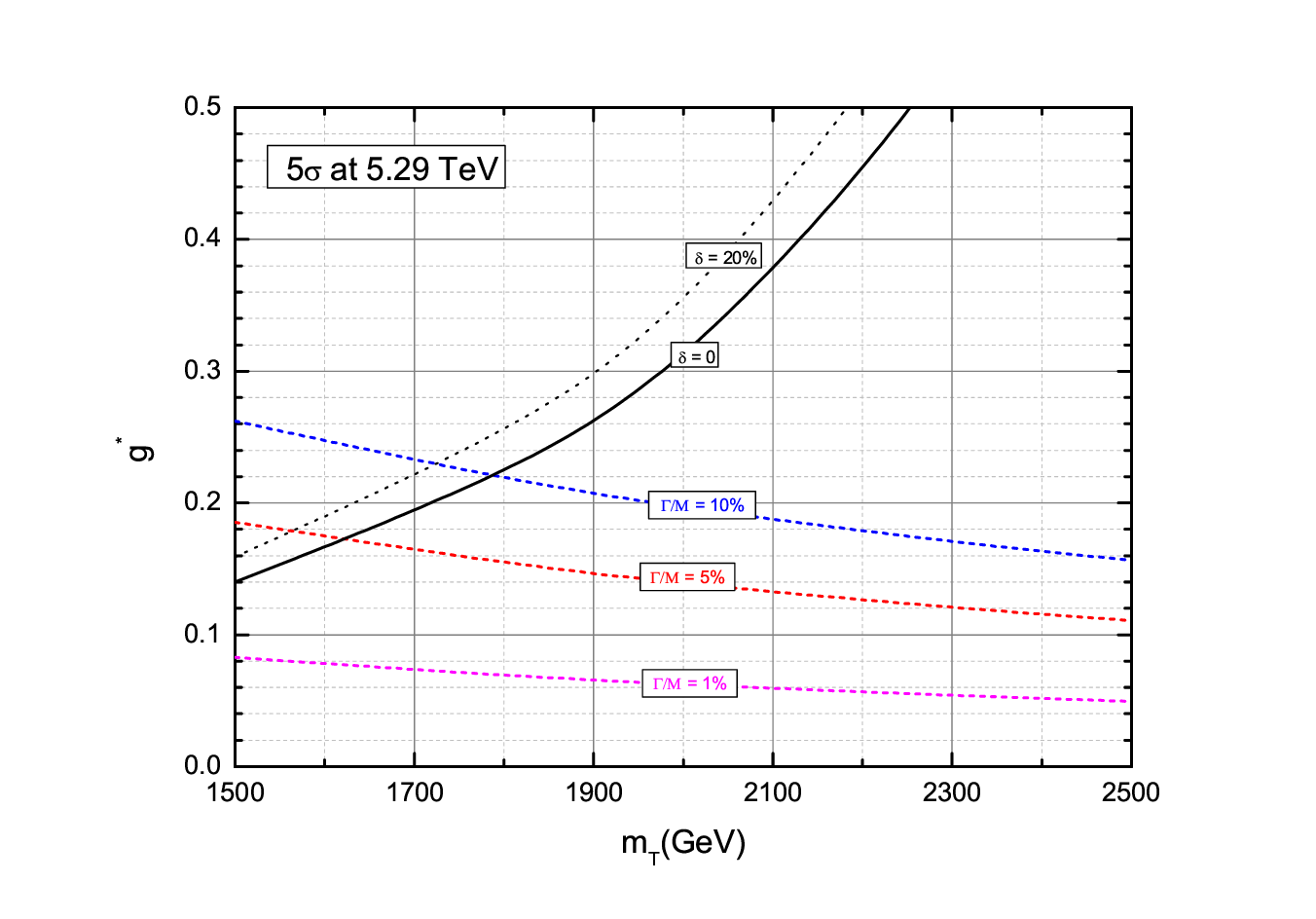}}
\centerline{\epsfxsize=7.5cm \epsffile{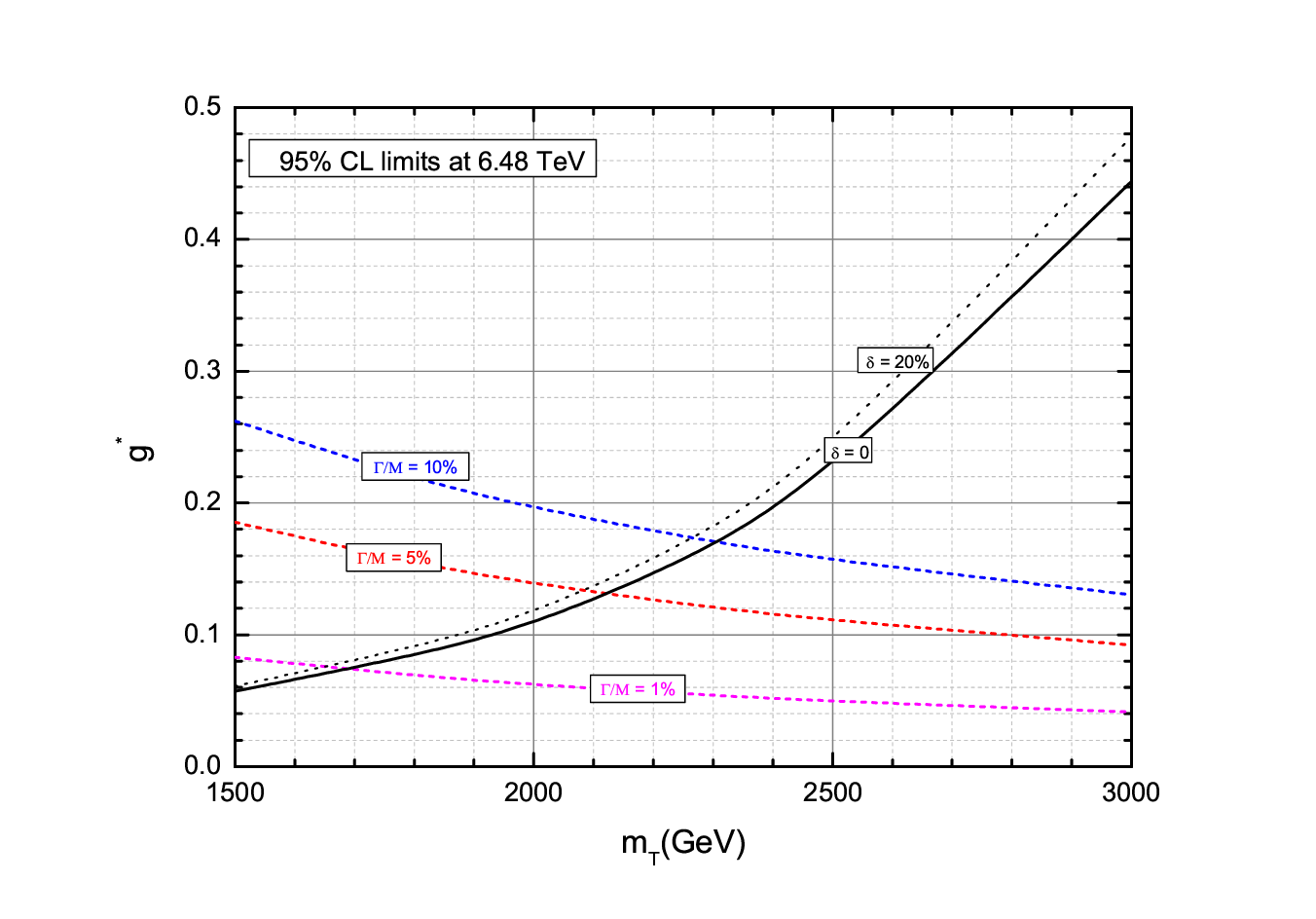}\epsfxsize=7.5cm \epsffile{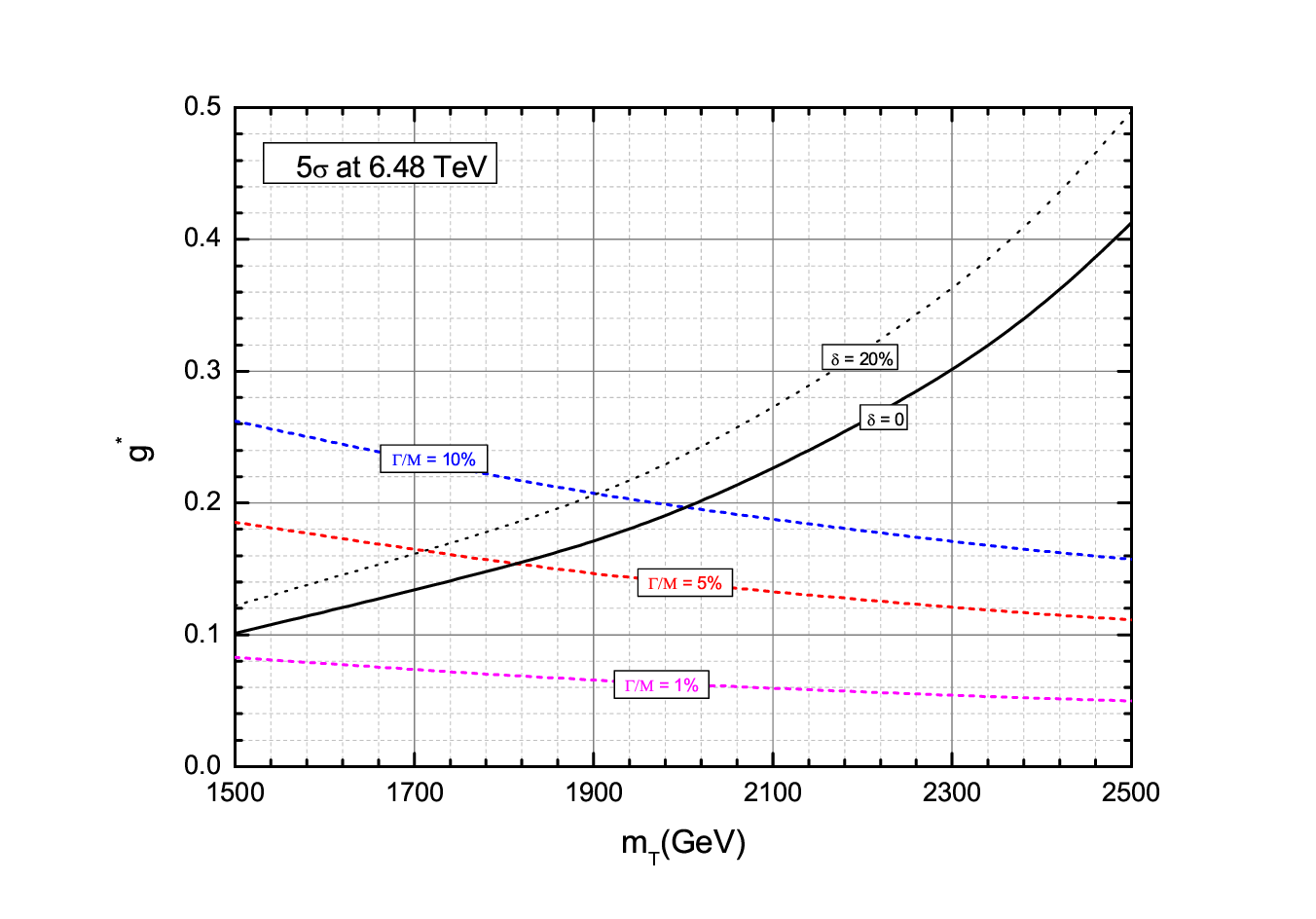}}
\centerline{\epsfxsize=7.5cm \epsffile{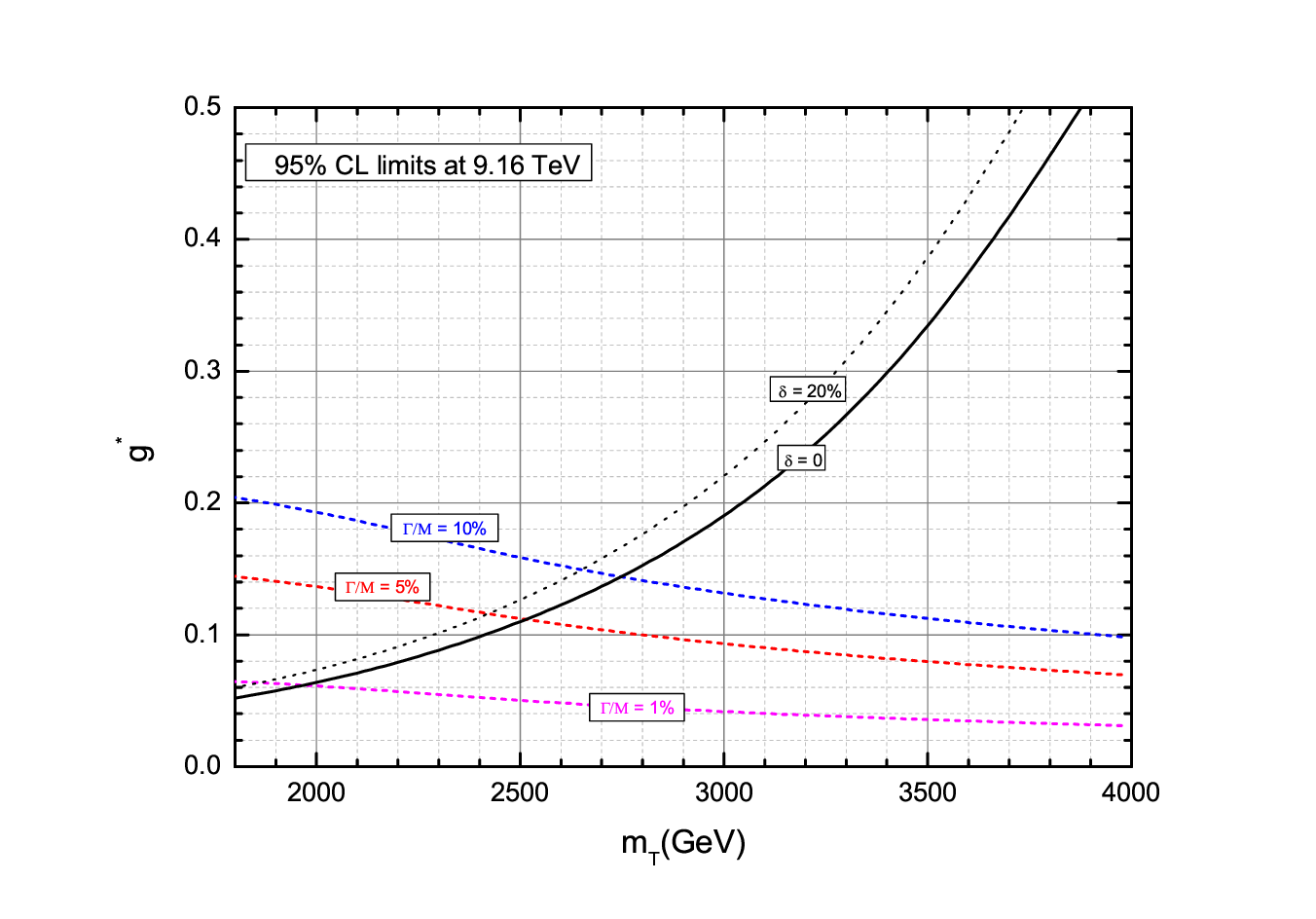}\epsfxsize=7.5cm \epsffile{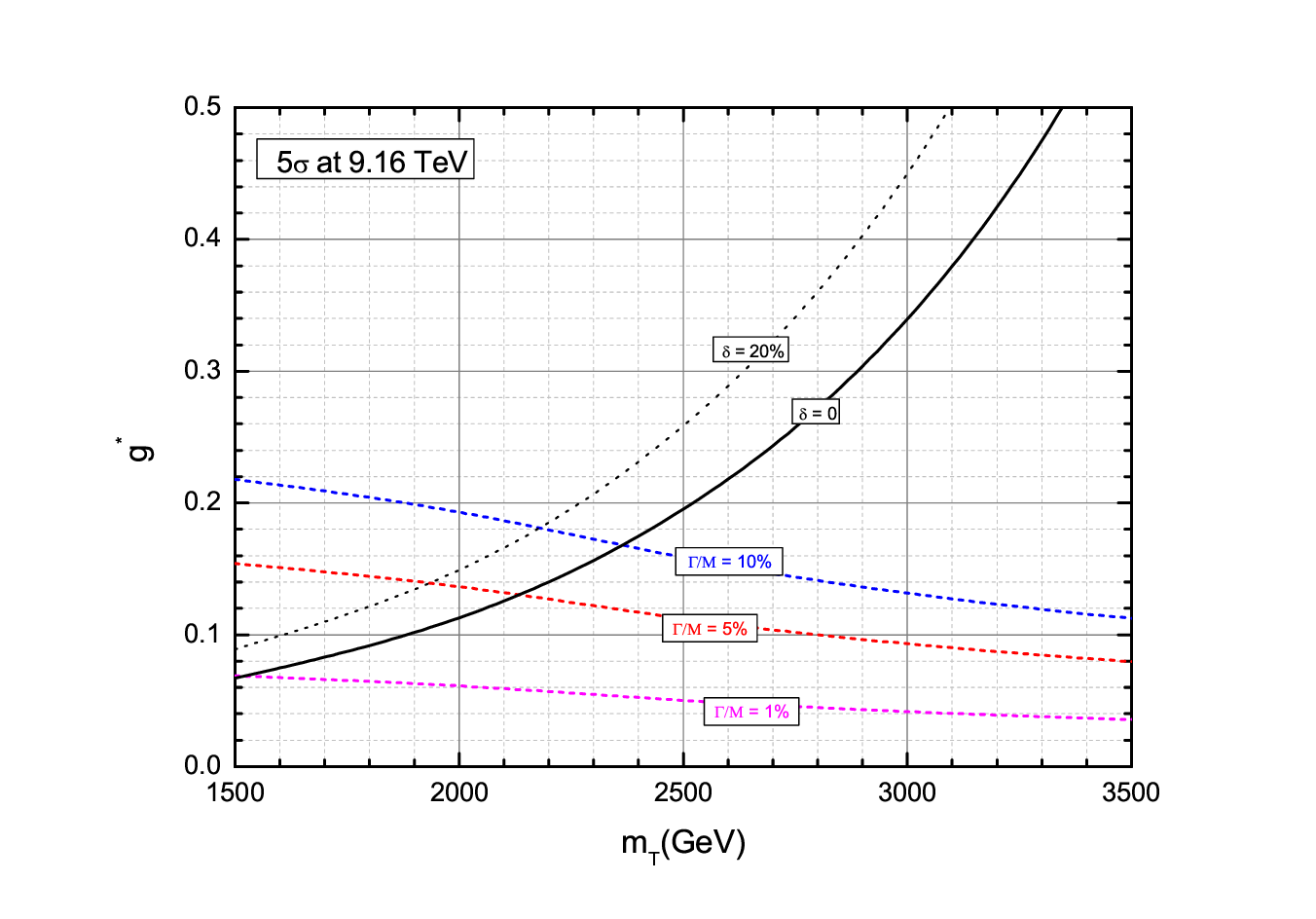}}
\caption{95\% CL exclusion limit (left panel) and $5\sigma$ discovery reach (right panel) contour plots for the signal in $g^{*}-m_T$ plane at $\sqrt{s}=5.29$, $6.48$, and $9.16$ TeV with $1000~\text{fb}^{-1}$ and $\delta_{sys}=0, 20\%$. Dashed lines indicate $\Gamma_T/m_T = 1\%, 5\%, 10\%$.}
\label{fig4}
\end{center}
\end{figure}

Figure~\ref{fig4} presents the $95\%$ CL exclusion limits and $5\sigma$ discovery reaches in the $g^{*}$-$m_T$ plane for the three c.m. energies. To assess the impact of systematic uncertainties on the significances, we consider two scenarios: no systematic uncertainty ($\delta_{sys}=0$) and a typical systematic uncertainty of $\delta_{sys}=20\%$. As expected, a larger systematic uncertainty degrades both the discovery potential and the exclusion power.
For $\delta_{sys}=20\%$, the $5\sigma$ discovery regions are as follows: at $\sqrt{s}=5.29$~TeV, $g^{*}\in [0.16,0.5]$ for $m_T\in [1500,2200]$~GeV; at $\sqrt{s}=6.48$~TeV, $g^{*}\in [0.12,0.5]$ for $m_T\in [1500,2500]$~GeV; and at $\sqrt{s}=9.16$~TeV, $g^{*}\in [0.09,0.5]$ for $m_T\in [1500,3100]$~GeV. The corresponding $95\%$ CL exclusion regions are $g^{*}\in [0.08,0.44]$ for $m_T\in [1500,2500]$~GeV at $5.29$~TeV, $g^{*}\in [0.06,0.48]$ for $m_T\in [1500,3000]$~GeV at $6.48$~TeV, and $g^{*}\in [0.06,0.5]$ for $m_T\in [1500,3700]$~GeV at $9.16$~TeV.

So far, we have treated the VLQ-$T$ signals within the narrow-width approximation (NWA).
In this approximation, the total width $\Gamma_T$ is computed at leading order as the sum of the partial widths for the $T \to bW, tZ, th$ decays, derived from the Lagrangian in Eq. (1). This is valid when the width-to-mass ratio $\Gamma_T / m_T$ is at the per-mille level. However, as demonstrated in Refs.~\cite{Banerjee:2024zvg,Barducci:2013zaa}, when $\Gamma_T / m_T$ reaches the percent level or above, a full Breit-Wigner prescription should be adopted. To quantify the impact of finite-width effects on our sensitivity projections, we overlay in Fig.~4 the constant contours of $\Gamma_T/m_T$ for three representative values: $1\%$, $5\%$, and $10\%$.
 For a fixed ratio $\Gamma_T/m_T = 10\%$ and a systematic uncertainty of $20\%$, the $5\sigma$ discovery reach for the VLQ-$T$ mass is approximately $1700$, $1900$, and $2200$~GeV at $\sqrt{s}=5.29$, $6.48$, and $9.16$~TeV, respectively, with corresponding $95\%$ CL exclusion lower limits of about $2050$, $2280$, and $2650$~GeV at the three c.m. energies.

 We also note that theoretical cross-section calculations with finite-width effects and non-resonant contributions are only reliable for relative $T$-quark widths up to $\Gamma_T/m_T \sim 50\%$ (see, e.g., Ref.~\cite{Deandrea:2021vje}). Our results are therefore presented only within this restricted regime. A complete off-shell calculation including interference effects for the full parameter space will be pursued in our future work.

\begin{figure}[htb]
\begin{center}
\vspace{1.5cm}
\centerline{\epsfxsize=10cm \epsffile{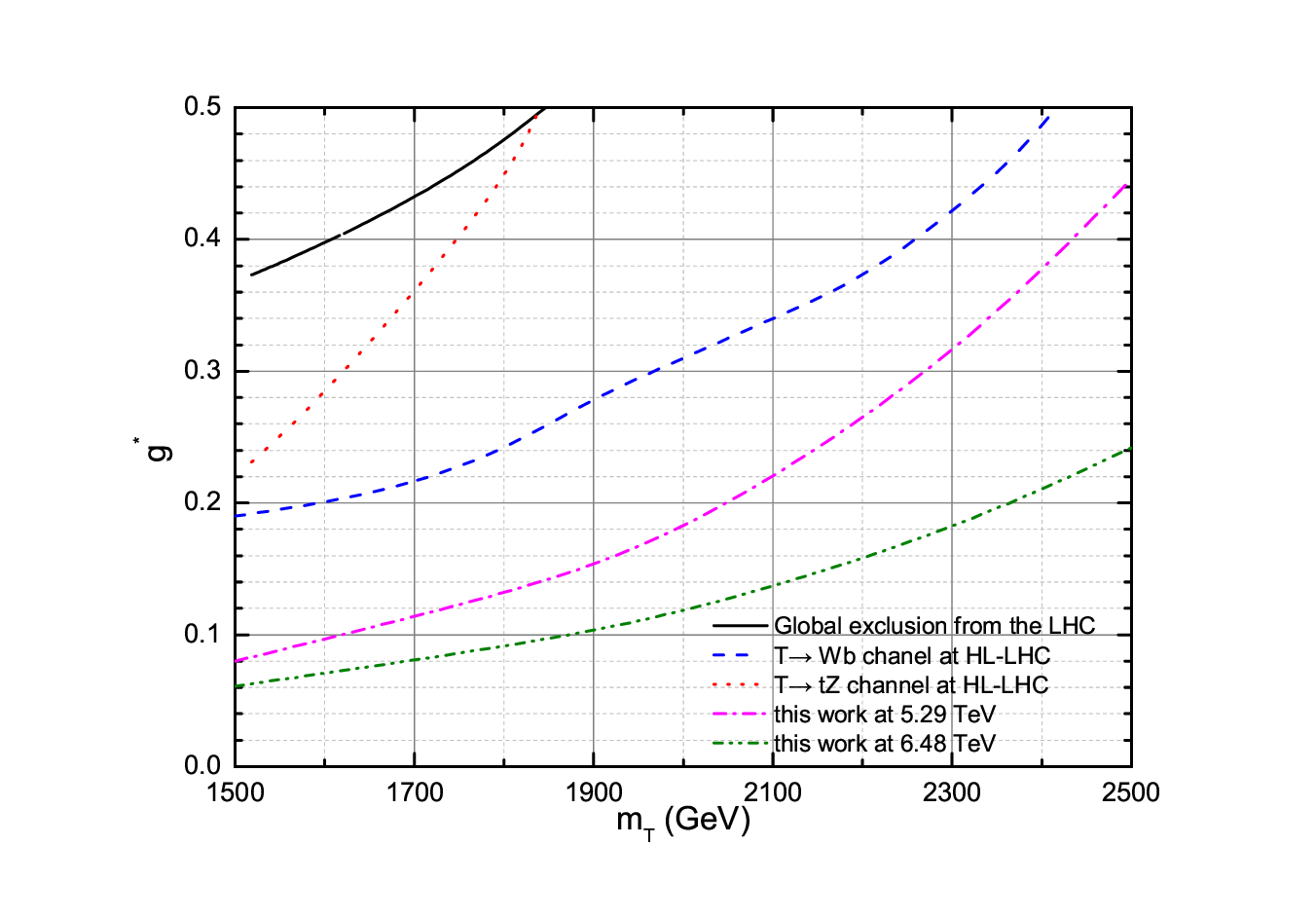}}
\caption{Comparison of the 95\% CL exclusion limits on the singlet VLQ-$T$ in the $(g^*, m_T)$ plane between current LHC bounds~\cite{Benbrik:2024fku}, projected HL-LHC sensitivities at $\sqrt{s}=14$~TeV with $3~\text{ab}^{-1}$~\cite{Yang:2021btv,Han:2023ied}, and our results for two representative $\mu p$ collider energies: $\sqrt{s} = 5.29$~TeV and $6.48$~TeV, with an integrated luminosity of $1000~\text{fb}^{-1}$ and a systematic uncertainty of $20\%$.}
\label{fig5}
\end{center}
\end{figure}

To facilitate a direct comparison of the VLQ-$T$ search capabilities across different collider scenarios, we summarize in Table~\ref{list} the exclusion and discovery reaches in the $(g^*, m_T)$ parameter space, and present in Fig.~\ref{fig5} a visual comparison of the $95\%$ CL exclusion limits among current LHC bounds~(based on the combined ATLAS and CMS exclusion limits for the singlet VLT scenario from Ref.~\cite{Benbrik:2024fku}), projected HL-LHC sensitivities~\cite{Yang:2021btv,Han:2023ied}, and our results for two representative $\mu p$ collider energies: $\sqrt{s}=5.29$~TeV and $6.48$~TeV. For the singlet VLQ-$T$ scenario, the mixing parameter $\kappa$ used in the LHC analyses of Ref.~\cite{Benbrik:2024fku} is related to the effective coupling $g^*$ adopted in this work by $g^* = \sqrt{2}\,\kappa$. After applying this conversion, the current LHC bounds exclude $g^* \gtrsim 0.37$ at $m_T \sim 1.5$~TeV, while our $\mu p$ collider analysis can probe $g^*$ values as low as $0.06$--$0.08$ at the same mass, demonstrating a significant improvement in sensitivity. Our analysis assumes a conservative systematic uncertainty of $20\%$, while the other studies adopt varying assumptions (see the corresponding references for details). As clearly demonstrated in both the table and the figure, a future $\mu p$ collider would provide significantly enhanced and complementary sensitivity for singlet VLQ-$T$ searches, particularly in the high-mass region, benefiting from a cleaner collision environment and substantially reduced QCD backgrounds compared to the LHC and HL-LHC.

\begin{table}[htbp]
\begin{center}
 \caption{\label{list}
Some results of searching for the singlet VLQ-$T$ at different high-energy colliders. Here, the symbol ``$\setminus$" stands for no relevant results in the reference. The results in this work correspond to a mild systematic uncertainty of 20\% at  a $\mu p$ collider with an integrated luminosity of $1000~\text{fb}^{-1}$.}
\vspace{0.2cm}
\begin{tabular}{c|c|cc|cc|c}
\hline\hline
\multirow{2}{*}{Channel}      &\multirow{2}{*}{Data Set} &\multicolumn{2}{c|}{Excluding capability} & \multicolumn{2}{c|}{Discovery capability}&\multirow{2}{*}{Reference}  \\
\cline{3-4} \cline{5-6}
&&$g^{\ast}$&$m_T/\rm TeV$&$g^{\ast}$&$m_T/\rm TeV$&  \\
\hline
$T\to tZ~(Z\to \ell^{+}\ell^{-})$&LHC @14 TeV, 3 ab$^{-1}$&[0.06, 0.25] & [0.9, 1.5]&[0.10, 0.42]&[0.9, 1.5]&\cite{Liu:2017sdg}  \\
$T\to th$&LHC @14 TeV, 3 ab$^{-1}$&[0.16, 0.5] & [1.0, 1.6]&[0.24, 0.72]&[1.0, 1.6]&\cite{Liu:2019jgp}  \\
$T\to bW^{+}$&LHC @14 TeV, 3 ab$^{-1}$&[0.19, 0.5] & [1.3, 2.4]&[0.31, 0.5]&[1.3, 1.9]&\cite{Yang:2021btv}  \\
$T\to tZ~(Z\to \nu\bar{\nu})$&LHC @14 TeV, 3 ab$^{-1}$&[0.09, 0.5] & [1.0, 1.84]&[0.14, 0.5]&[1.0, 1.6]&\cite{Han:2023ied}  \\
$T\to bW^{+}$&$e\gamma$ collider @2 TeV, 1 ab$^{-1}$&[0.13, 0.5] & [0.8, 1.6]&$\setminus$&$\backslash$&\cite{Yang:2018fcx}  \\
$T\to tZ$&$e\gamma$ collider @3 TeV, 3 ab$^{-1}$&[0.15, 0.23] & [1.3, 2.0]&[0.23, 0.5] &[1.3, 2.0]&\cite{Shang:2019zhh}  \\
$T\to th$&$e\gamma$ collider @3 TeV, 3 ab$^{-1}$&[0.14, 0.50] & [1.3, 2.0]&[0.27, 0.5] &[1.3, 2.0]&\cite{Shang:2020clm}  \\
$T\to bW^{+}$&$e^{+}e^{-}$ collider @3 TeV, 5 ab$^{-1}$&[0.15, 0.40] & [1.5, 2.6]&[0.24, 0.44] &[1.5, 2.4]&\cite{Qin:2022mru}  \\
 $T\to tZ$&$e^{+}e^{-}$ collider @3 TeV, 5 ab$^{-1}$&[0.19, 0.40] & [1.3, 2.5]&[0.31, 0.5] &[1.3, 2.3]&\cite{Han:2022exz} \\  \hline
\multirow{3}{*}{$T\to tZ\to b\ell \nu J$}&$\mu p$ collider @5.29 TeV&[0.08, 0.44] & [1.5, 2.5]&[0.16, 0.5] &[1.5, 2.2]&\multirow{3}{*}{This work}  \\
&$\mu p$ collider @6.48 TeV&[0.06, 0.48] & [1.5, 3.0]&[0.12, 0.5] &[1.5, 2.5]& \\
&$\mu p$ collider @9.16 TeV&[0.09, 0.4] & [1.5, 3.7]&[0.09, 0.5] &[1.5, 3.1]&  \\
\hline
 \end{tabular}
 \end{center}
\end{table}

It is instructive to compare the sensitivities obtained in this work with those from the recent study of the $T\to Wb$ channel at the same $\mu p$ collider~\cite{Han:2025itd}. That analysis, performed with $100~\text{fb}^{-1}$ and $10\%$ systematic uncertainty, reported $5\sigma$ discovery reaches of $m_T = 2520$, $2900$, and $3750$~GeV at $\sqrt{s}=5.29$, $6.48$, and $9.16$~TeV, respectively, assuming $g^*=0.5$. In the present work, with $1000~\text{fb}^{-1}$ but a more conservative $20\%$ systematic uncertainty, we obtain $5\sigma$ discovery limits of $m_T = 2200$, $2500$, and $3100$~GeV at the three c.m. energies for the same coupling, with corresponding $95\%$ CL exclusion limits of $m_T = 2500$, $3000$, and $3700$~GeV. Although the $T\to tZ$ channel has a branching ratio roughly half that of $T\to Wb$, the two channels are complementary beyond their branching fractions. The $T \to tZ$ signature features a boosted fat jet from the $Z$ boson, whereas the $T \to Wb$ channel relies on a fat jet from the $W$ boson. These different final-state topologies imply distinct sensitivities to jet energy scale uncertainties and QCD background modeling. A combined analysis of both decay modes would therefore not only increase the overall signal yield but also provide a robust cross-check against systematic biases, thereby strengthening the case for a potential discovery at a future $\mu p$ collider.
\section{Conclusions and Discussion}

In this paper, we have investigated the discovery potential of a future $\mu p$ collider for a heavy singlet VLQ-$T$ produced via the single-production mode $\mu g \to \nu_\mu b T$, followed by the decay chain $T \to tZ$, with $t \to bW$ (leptonic $W$ decay) and $Z \to q\bar{q}$ (reconstructed as a single fat jet). Within a simplified two-parameter framework defined by $m_T$ and $g^*$, we have performed a detailed detector-level simulation for three center-of-mass energies: $\sqrt{s} = 5.29$, $6.48$, and $9.16$~TeV. After optimizing the event selection criteria, we have derived the $5\sigma$ discovery and $95\%$ CL exclusion limits in the $(g^*, m_T)$ parameter space.

Assuming an integrated luminosity of $1000~\text{fb}^{-1}$ and a systematic uncertainty of $20\%$, the $5\sigma$ discovery regions are as follows: at $\sqrt{s}=5.29$~TeV, $g^* \in [0.16, 0.5]$ for $m_T \in [1500, 2200]$~GeV; at $\sqrt{s}=6.48$~TeV, $g^* \in [0.12, 0.5]$ for $m_T \in [1500, 2500]$~GeV; and at $\sqrt{s}=9.16$~TeV, $g^* \in [0.09, 0.5]$ for $m_T \in [1500, 3100]$~GeV. The corresponding $95\%$ CL exclusion regions are $g^* \in [0.08, 0.44]$ for $m_T \in [1500, 2500]$~GeV at $5.29$~TeV, $g^* \in [0.06, 0.48]$ for $m_T \in [1500, 3000]$~GeV at $6.48$~TeV, and $g^* \in [0.06, 0.5]$ for $m_T \in [1500, 3700]$~GeV at $9.16$~TeV. These sensitivities are competitive with, and complementary to, those obtained from the $T \to Wb$ channel as well as from other collider platforms.

For completeness, we briefly comment on three alternative final-state topologies of the $T \to tZ$ cascade that are not analyzed in this work, in contrast to our baseline channel with a semileptonic top decay and a hadronic $Z$ decay. The omitted signal modes are:
\begin{itemize}
    \item invisible $Z \to \nu\bar{\nu}$ decay with fully hadronic top decay $t \to bW \to jjb$;
    \item invisible $Z \to \nu\bar{\nu}$ decay with semileptonic top decay $t \to bW \to b\ell\nu$;
    \item leptonic $Z \to \ell^+\ell^-$ decay with fully hadronic top decay $t \to bW \to jjb$.
\end{itemize}
Preliminary MC estimations indicate that all three alternative channels suffer from two critical drawbacks compared to our baseline. First, their effective signal cross sections are noticeably smaller, yielding far fewer signal events under the same integrated luminosity. Second, they are subject to substantially larger SM backgrounds. In particular, the two invisible $Z$ channels are overwhelmed by irreducible backgrounds from single-top and multi-jet QCD production, which produce large missing transverse momentum and cannot be efficiently suppressed by simple kinematic cuts. Meanwhile, the channel with fully hadronic top decay and leptonic $Z$ decay loses the powerful background rejection provided by an isolated high-$p_T$ charged lepton from the semileptonic $W$ decay, and thus suffers from severe multi-jet contamination. Given the limited signal statistics and intractable background levels, we do not carry out full detector simulations or significance calculations for these channels. More sophisticated multivariate discriminants or advanced jet substructure tagging algorithms may partially improve their discovery potential, and could be investigated in dedicated follow-up studies.

Finally, we note that our cross section calculations are performed at leading order with a single PDF set (NNPDF23LO1). A complete next-to-leading order (NLO) analysis, including the corresponding K-factors and PDF uncertainties, is beyond the scope of this work. However, such higher-order corrections are not expected to qualitatively alter the main conclusions of our study. The dominant effect of NLO corrections is typically a moderate shift in the overall normalization, while the shape of the kinematic distributions and the relative signal-to-background ratios are generally stable against such corrections. Moreover, PDF uncertainties are expected to be subdominant compared to the $20\%$ systematic uncertainty adopted in our significance estimates. Therefore, our sensitivity projections remain robust as a conservative benchmark for the discovery potential at a future $\mu p$ collider.
\begin{acknowledgments}
This work of Y.-B. Liu is supported by the Natural Science Foundation of Henan Province~(Grant No.~252300421988).
\end{acknowledgments}


\end{document}